\documentclass{article}
\usepackage{spconf,amsmath,graphicx,hyperref}
\usepackage{amsfonts}
\usepackage[numbers]{natbib}
\usepackage{booktabs}
\usepackage{makecell}
\usepackage{xspace}

\title{Multi-layer attentive probing improves transfer of audio representations for bioacoustics}
\name{{\small\shortstack{Marius Miron, David Robinson, Masato Hagiwara, Titouan Parcollet, Jules Cauzinille, Gagan Narula\\Milad Alizadeh, Ellen Gilsenan-McMahon, Sara Keen, Emmanuel Chemla, Benjamin Hoffman\\Maddie Cusimano, Diane Kim, Felix Effenberger, Jane K. Lawton, Aza Raskin\\Olivier Pietquin, Matthieu Geist}}}
\address{Earth Species Project}
\begin{document}
\ninept

\newcommand\beans{BEANS\xspace}
\newcommand\beanscls{BCLS\xspace}
\newcommand\dog{DOG\xspace}
\newcommand\bat{BAT\xspace}
\newcommand\msq{HBDB\xspace}
\newcommand\cbi{CBI\xspace}
\newcommand\bwtk{BWTK\xspace}
\newcommand\beansdet{BDET\xspace}
\newcommand\ena{ENA\xspace}
\newcommand\rfcx{RFCX\xspace}
\newcommand\hic{HIC\xspace}
\newcommand\gib{GIB\xspace}
\newcommand\dcase{DCASE\xspace}

\newcommand\bset{BirdSet\xspace}
\newcommand\pow{POW\xspace}
\newcommand\per{PER\xspace}
\newcommand\nbp{NBP\xspace}
\newcommand\hsn{HSN\xspace}
\newcommand\uhh{UHH\xspace}
\newcommand\sne{SNE\xspace}
\newcommand\ssw{SSW\xspace}
\newcommand\nes{NES\xspace}

\newcommand\inid{ID\xspace}
\newcommand\pip{PIP\xspace}
\newcommand\chif{CHIF\xspace}
\newcommand\mac{MAC\xspace}
\newcommand\owl{OWL\xspace}
\maketitle
\begin{abstract}
Probing heads map the representations learned from audio by a machine learning model to downstream task labels and are a key component in evaluating representation learning. Most bioacoustic benchmarks use a fixed, low-capacity probe, such as a linear layer on the final encoder layer. While this standardization enables model comparisons, it may bias results by overlooking the interaction between encoder features and probe design. In this work, we systematically study different probing strategies across two bioacoustic benchmarks, BEANs and BirdSet. We evaluate last- and multi-layer probing, across linear and attention probes. We show that larger probe heads that leverage time information have superior performance. Our results suggest that current benchmarks may misrepresent encoder quality when relying on a last-layer probing setup. Multi-layer probing improves downstream task performance across all tested models, while attention probing has superior performance to linear probing for transformer models.   
\end{abstract}
\section{Introduction}
\label{sec:intro}
\begin{figure*}[h]
\centering
  \includegraphics[width=1.\textwidth]{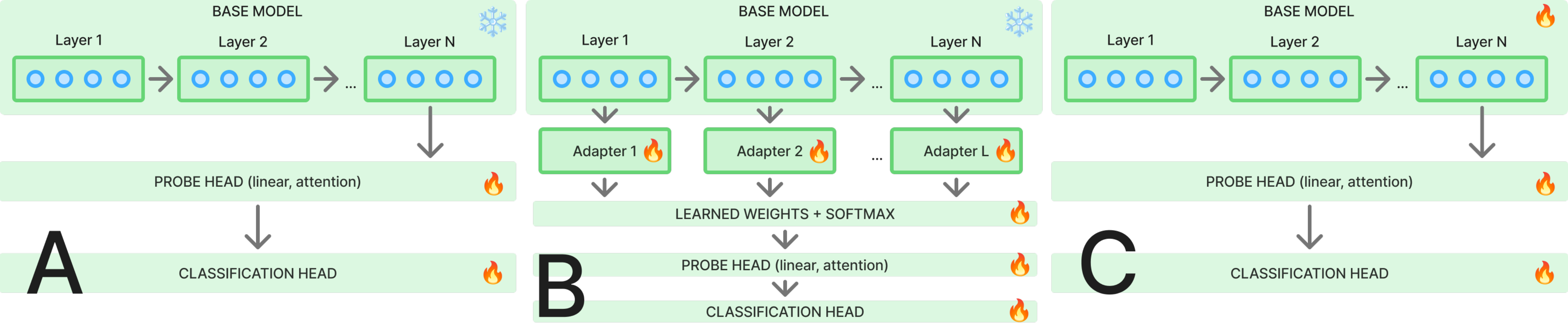}
  \caption{The architecture of the probes we use for a given base model. We test three main probing strategies: \textbf{A.} Last-layer probing and \textbf{B.} All-layer probing \textbf{C.} Fully fine-tuning. The fire symbol denotes that the model is trainable and the snowflake that the parameters are frozen.}
  \label{fig:diagram}
\end{figure*}
Bioacoustics \citep{bradbury1998principles} is an interdisciplinary field that examines the production and perception of sound in animals, linking acoustic signals to behavioral, ecological, and evolutionary processes.
Recently, the field benefited from the use of deep learning \citep{stowell2022computational}, either by developing domain-specific approaches or by tailoring an existing audio encoder for a given dataset.
Bioacoustic encoders learn representations from animal sounds for tasks such as species classification \citep{hagiwara2023beans}, detection \citep{kahl2021birdnet,van2025perch,rauchbirdset}, and individual identification \citep{stowell2019automatic}, supporting applications in biodiversity monitoring \citep{stowell2022computational} or animal communication \citep{bradbury1998principles}.

Recent work \citep{van2025perch, miron2025matters} has explored what matters for training such encoders, evaluating the impact of training data diversity, model architectures, and learning paradigms across large-scale benchmarks such as BEANS~\citep{hagiwara2023beans} and BirdSet~\citep{rauchbirdset}. These studies have produced versatile encoders applicable to species classification, individual identification, and behavioral analysis, yet they share a common limitation: evaluation typically relies on fixed, low-capacity neural networks on top of the final encoder, commonly known as {\it probing heads}.
In speech representation learning, this practice has been questioned. Studies have shown that encoder rankings and performance can change dramatically when altering the probing head architecture~\citep{yang2021superb, tsai2022superb, shi2023ml, shi2024ml}, suggesting that current benchmarks may conflate encoder quality with probe design \citep{zaiem2025speech}. Low-capacity probes may fail to exploit information spread across layers or time. As a result, benchmarks that fix the probe design risk underestimating some models and overestimating others.

We address this issue for bioacoustics by benchmarking multiple audio encoders usage, probing strategies, and probing heads. We focus here on single- and multi-layer probing, comprising adapter modules for aligning layer outputs, and learned layer weights. We evaluate two probe types—linear and attention-based probes. In addition, we evaluate fully fine-tuning the model with the respective probe heads attached to the last layer as a top-line. %

Our experiments on BEANS and BirdSet benchmarks show that probe choice significantly affects encoder performance with multi-layer probing giving better performance than last-layer probing, and attention probing being particularly useful when using embeddings extracted from self-supervised learning models. We propose concrete recommendations for practitioners when opting for the right probing strategy and head, given a task and dataset. More, we advocate for introducing larger probe heads and different probing strategies in bioacoustics benchmarks. We open source the code as a Python library \citep{avex}.

\section{RELATION TO PRIOR WORK}
\label{sec:prior}
In this paper we build upon the existing bioacoustics benchmarks and extend them in a four folded manner.
Layer-wise probing of speech models has been studied in the context of bioacoustics benchmarks \cite{cauzinille2025crossing,sarkar2025comparing}. In contrast to these we propose to \textbf{(1) leverage embeddings extracted from all the blocks of the base models}, a de-facto strategy for speech benchmarks such as SUPERB \citep{tsai2022superb,zaiem2025speech}. 
Adding to that, we average the layers using a set of \textbf{(2) interpretable learned weights} that we can explored for each dataset.
However, in contrast to the SUPERB setup we aim at evaluating architectures besides transformers, potentially aggregating embeddings of different shapes. Importantly, a wide choice as off-the-shelf models, EfficientNets such as BirdNET\citep{kahl2021birdnet} are widely used and linear probing on the last layer is the de-facto strategy. Thus, \textbf{(3) we introduce adapters as part of the probing strategy} to project embeddings of different shapes to the same dimension. 
In contrast to the previous work on evaluating foundation models for bioacoustics ~\citep{ghani2023global, hamer2023birbgeneralizationbenchmarkinformation, van2025perch, miron2025matters} \textbf{(4) we provide an extensive probing study to disentangle} the base model types (self-supervised learning, SSL vs supervised learning, SL), architectures (transformer vs CNNs), probing strategies (multi- vs last-layer), and heads (attention vs linear), towards having a comprehensive evaluation of the popular design choices when transferring representations for downstream bioacoustics tasks. 

\section{METHODS}
\label{sec:methods}

The evaluation framework including fully fine-tuning and the two main probing strategies, last- and multi-layer, are presented in Figure \ref{fig:diagram} and defined in Section \ref{ssec:task_definition}. We introduce the probe and classification heads in Section \ref{ssec:probe}, and the base models in Section \ref{ssec:base_models}. We give details about the datasets, tasks, evaluation metrics, hyperparameters in Section \ref{ssec:setup}.

\subsection{Task definition: probe strategies}
\label{ssec:task_definition}

Let $f_{\theta}$ be a pre-trained base model with parameters $\theta$.
For a base model with $L$ layers, let $h^{l} \in \mathbb{R^*}$ denote the hidden representation i.e the embeddings at layer $l$, with $*$ representing variable dimensions that depend on the type of layer, or more precisely $\mathbb{R}^{d_1 \times d_2 \times d_3}$ for convolutional layers and $\mathbb{R}^{d_1 \times d_2}$ for attention or recurrent layers.  
Given a downstream task and dataset $\mathcal{D} = \{(x_i, y_i)\}_{i=1}^N$ with $x_i \in \mathcal{X}$ audio files and $y_i \in \mathcal{Y}$ the labels, we define different probing strategies to extract and utilize embeddings from the base model.

\noindent\textbf{A. Last-layer probing.}
The base model parameters $\theta$ remain frozen, and only the probe parameters $\phi$ are optimized: $\hat{y} = g_{\phi}(h^L)$. 

\noindent\textbf{B. All-layer probing.}
Because choice of the layer may impact the performance on the downstream task, we extract representations from all blocks of the base models. We aim at aggregating heterogeneous layer representations resulting in embeddings of different dimensions. Thus, we need to project the embeddings to a common dimension so that the probe may work on them.
We design adapters $A_{\psi_l}: \mathbb{R}^* \rightarrow \mathbb{R}^{T_{\max} \times F_{\max}}$ to project a layer $l$ to a unified two-dimensional format: $(T_{\max}, F_{\max})$, where $T_{\max}$ is the maximum sequence length across all embeddings, and $F_{\max}$ is the maximum feature dimension. Each adapter outputs $\hat{h}^{l} = A_{\psi_l}(h^{(l)})$ with $ \tilde{h}^{l} \in \mathbb{R}^{T_{\max} \times F_{\max}}$. Importantly, we design adapters to process time and feature dimension separately. Note that for a CNN layer activations with dimensions $(d_1,d_2,d_3)$ as (channel, height, width), the time dimension corresponds to the width of the activations $d_3$ given that these models take as input spectrograms, and the feature to the product of the channels and height $d_2 \times d_4$.

The adapter is a two step operation comprising 
a linear projection from feature dimension $d_2$ to $F_{\max}$ such as $\tilde{h}^{(l)} = A_{\psi_l}^1(h^{(l)}): \mathbb{R}^{d_1 \times d_2} \rightarrow \mathbb{R}^{d_1 \times F_{\max}}
$
and an interpolation from sequence length $d_1$ to $T_{\max}$ such as $\hat{h}^{(l)} = A_{\psi_l}^2(\tilde{h}^{(l)}): \mathbb{R}^{d_1 \times F_{\max}} \rightarrow \mathbb{R}^{T_{\max} \times F_{\max}}$.

We then compute the weighted sum of the embeddings $h = \sum_{l=1}^L \alpha_l \hat{h}^{(l)}$, where $\alpha_l = \frac{\exp(w_l)}{\sum_{k=1}^L \exp(w_k)}$ are softmax-normalized weights derived from learnable parameters $\mathbf{w} = [w_1, w_2, \ldots, w_L]^\top \in \mathbb{R}^L$, ensuring $\sum_{l=1}^L \alpha_l = 1$. This allows us to interpret which layers are useful for each downstream task, dataset, and base model. 

\noindent\textbf{C. Fully fine-tuning.} 
All the parameters of the base model model $\theta$ are updated during training along with a probe head $g_{\phi}$, $\hat{y} = g_{\phi}(h^L)$ attached to the last layer, where $h^L = f_{\theta}$ are the embeddings from the last layer.

\subsection{Probe and classification heads}
\label{ssec:probe}

We compare two types of probing: linear and  attention-based. 
The linear probe first averages encoder representations over the time dimension, producing a single feature vector per input. It keeps the full feature dimension and applies a single linear layer to predict class scores. This design captures global information while keeping probe capacity low.
As a higher capacity alternative we chose an attention probe that models time dependencies between features. The attention-based probe assigns learnable weights to each time step using soft attention, enabling the model to focus on informative regions. The weighted features pass through a residual connection, followed by layer normalization and dropout for regularization. A second dropout layer is applied before the final classification layer. This architecture increases probe capacity while preserving temporal structure in the representations. The fully fine-tuned models add to the last layer probe parameters a large number of parameters from the unfrozen base model.

\begin{table*}[ht]
\centering
\caption{Base models included in the probe experiments. We compare self-supervised (SSL) and supervised (SL) models across four different probe experiments: two probe configurations, A. last layer probing and B. all layers probing, and two probe types, linear and attention. Fully fine-tuning is a more expensive setup, training base model and last parameters.}
\label{tab:base_models}
\begin{tabular}{llllllllll}
\toprule
\textbf{Model}           & \textbf{Pre-training Data}            & \textbf{Post-training Data} & \makecell{\textbf{N.P. base} \\ \textbf{(frozen)}} & \multicolumn{4}{c}{\textbf{N.P. probe (trainable)}} & \\
\cmidrule{5-8}
& & & & \multicolumn{2}{c}{\textbf{Attention}} & \multicolumn{2}{c}{\textbf{Linear}} & \\
\cmidrule{5-6} \cmidrule{7-8}
& & & & \textbf{Last} & \textbf{All} & \textbf{Last} & \textbf{All} & \\
\midrule
BEATs (SSL)                                       & AudioSet + Speech                     & --    & 90.35M & 2.40M & 2.40M & 37.68K & 1.23M &                                      \\
EAT (SSL)                                          & Bio  + AudioSet                   & --    & 90.01M & 2.40M & 2.40M & 37.68K & 414.00K &                                       \\
BirdAVES (SSL)                                        & Xeno Canto                      & --   & 94.37M & 2.40M & 2.40M & 37.68K & 414.00K &                                       \\
\midrule
BEATs (SL)                                        & AudioSet + Speech                     & Bio + AudioSet    & 90.35M & 2.40M & 2.40M & 37.68K & 1.23M &                                      \\
NatureBEATs (SL)                           & AudioSet + Speech                     & Bio + Text prompts   & 90.35M & 2.40M & 2.40M & 37.68K & 1.23M &                \\
EAT (SL)                 & Bio + Audioset   & Bio + AudioSet  & 90.01M & 2.40M & 2.40M & 37.68K & 414.00K &        \\
EfficientNet (SL)                              & ImageNet                               & Bio + AudioSet & 5.29M & 105.14M & 41.04M & 250.93K & 677.42K &      \\
\bottomrule
\end{tabular}
\end{table*}

\subsection{Base models}
\label{ssec:base_models}

We benchmark two main categories of base models: self-supervised and supervised encoders. All models were introduced in the previous "What Matters" study~\citep{miron2025matters} but were evaluated there only with a simple linear probe on the last layer embeddings. In this work, we keep the same set of models but test them with additional probing strategies and heads. We summarize the base models, their training data in Table \ref{tab:base_models}. We include the number of parameters for the base model, in our case frozen, and the number of parameters for the created probes. Note that when probing multiple layers of same dimension the adapters are not needed (transformers), whereas layers of very different shapes require large adapters because we always project to the highest dimension (CNNs). This may result into larger probes, such as the ones used for EfficientNet. 
For the models proposed in~\citep{miron2025matters}, we take the checkpoints trained on combined bioacoustics and general audio data which achieved average best performance across the benchmarks. 

\noindent\textbf{Self-supervised models.}
We include two general-purpose audio encoders: BEATs \citep{chen2023beats} and EAT \citep{chen2024eat}, both pre-trained on large audio or speech datasets using self-supervised objectives. BEATs is a state-of-the-art model trained on AudioSet and speech corpora, while EAT is fully open-source and designed for efficiency and flexible pre-training. From the bioacoustics domain, we include BirdAVES~\citep{hagiwara2023aves}, the checkpoint trained on birds (Xeno-Canto), a HuBERT-based model trained self-supervised on bird sounds.

\noindent\textbf{Supervised models.}
We include EfficientNet trained on bioacoustics and general audio with standard supervised learning. Because we found impossible to extract representations from inner layers of the tensorflow models (BirdNET~\citep{kahl2021birdnet}, Perch~\citep{van2025perch}) we did not include these models in this study, however they have a similar architecture to our EfficientNet model and they used similar training data. To compare supervised and self-supervised approaches within the same architecture, we also post-train BEATs and EAT with supervised objectives on bioacoustic data and general audio. In addition, we benchmark NatureBEATs \citep{robinsonnaturelm}, the post-trained BEATs encoder from the NatureLM-audio model.

\subsection{Evaluation setup}
\label{ssec:setup}

We evaluate the probing strategies applied to the base models on classification and detection tasks from the BEANS~\citep{hagiwara2023beans} and BirdSet~\citep{rauchbirdset} benchmarks. 
In terms of taxa and tasks, BirdSet focuses on bird species detection, considering a domain shift between train (cleaner recordings comprising one focal species) and test (noisy soundscapes with multiple species). For BirdSet, we follow the “Dedicated Train” setup, keeping separate train and test splits for each dataset. BEANS has a varied set of tasks and taxa comprising classification (bat individuals, bird species, dog breeds, marine mammal species, mosquito species) and detection tasks (birds, frogs and gibbons). For BEANS, we use the official splits. 

Single-class classification tasks, such as species or individual identification, use the standard cross-entropy loss, where each sample belongs to exactly one class. Multi-class classification tasks, such as sound event detection with overlapping events, use the binary cross-entropy (BCE) loss applied independently to each class, allowing multiple positive labels per sample. The downstream task training setup is similar to \citep{miron2025matters}. We train for $50$ epochs, instead of the $900$ used in \citep{miron2025matters} because running four time the number of experiments ran in \citep{miron2025matters} would be prohibitive in terms of costs. To reduce overfitting, we introduced a cosine learning rate scheduler with the first $5$ epochs being the learning stage. We use a learning rate of $0.0001$ and an AdamW optimizer, as BEANS and BirdSet.
In terms of evaluation metrics, we report top-1 accuracy on single-label tasks (classification) and macro-averaged mean average precision (mAP) for multi-label tasks (detection).

Because probing all possible layers is computationally expensive and all base model organize their layers in blocks, we extract embeddings from the last layer in each block, which are usually the `fc' or `fc2' layers for the transformers and the last convolution in each block for EfficientNet. We found that the performance of the EAT model significantly increased when using the output of the attention layer instead `attn\_out' and we used this layer for all the two EAT models. Given the total number of blocks in the base models  probing all layers corresponds to $11$ layer embeddings for BEATs and AVES, $10$ for EAT, and $15$ layer embeddings for EfficientNet. Note that when these embeddings have the same temporal and feature dimension, no adapters are used.

\section{RESULTS}
\label{sec:pagestyle}

\begin{figure}[h]
\centering
  \includegraphics[width=0.49\textwidth]{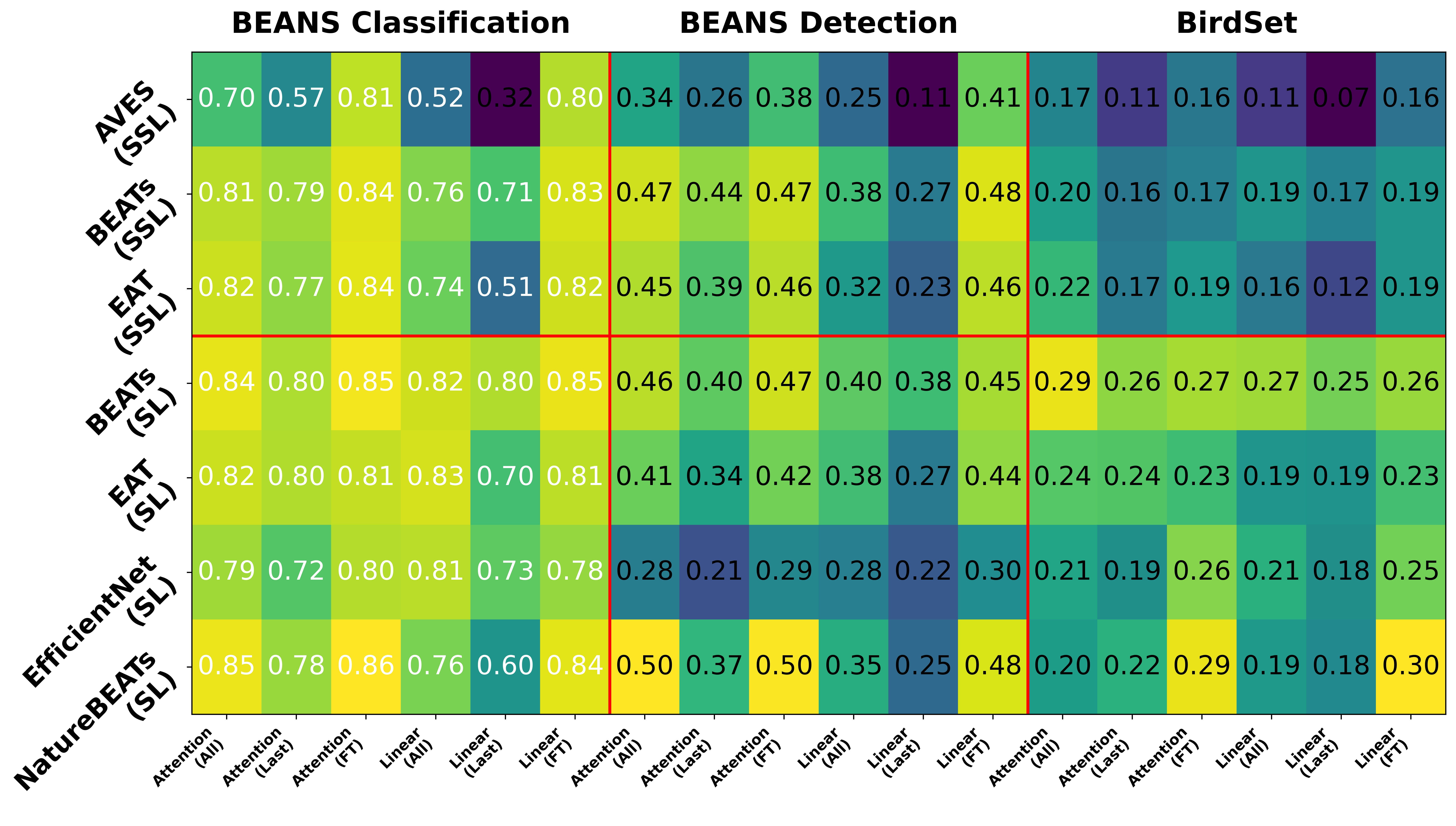}
  \caption{Benchmark-averaged results in terms of top-1 accuracy for BEANS classification and mean average precision for BEANS detection and BirdSET across the base models and the six pairs of probe configurations and heads. }
  \label{fig:results}
\end{figure}

Due to space constraints, we display the results averaged for each benchmark in Figure \ref{fig:results}. We split the results for BEANS into classification and detection.

As a direct comparison with the linear probe on last-layer experiments in \cite{miron2025matters}, we observe that our results are slightly lower for Linear (Last). This may happen because we train for less epochs and in an online fashion i.e. producing embeddings from the base model at each step rather than consuming them from the disk. These adjustments were made because compute and disk space would be prohibitive when considering multiple layers and a large number of epochs. 
Rather than obtaining a best model we aim at comparing different probing approaches under comparable conditions. 

\noindent\textbf{Multi-layer probe heads improve cross-taxa transfer.}
We observe a consistent trend for all base models and probe heads: extracting embeddings from all the blocks of the models leads to better performance than using solely the last layer embedding, which is the common probe strategy for bioacoustics evaluation \citep{rauchbirdset,rauch2025maskedautoencoderslistenbirds,van2025perch,miron2025matters}.
For all transformer models the gains when using `All' multi-layer probing are around $0.08$ (accuracy) for BEANS classification and detection and $0.03$ (mAP) for BirdSet, lower for EAT and larger for AVES, the former having been trained on birds recordings exclusively. 
To that extent, given that birds are predominant in the training data, we observe that the gains are less for BirdSet and more for non-birds benchmarks, particularly BEANS classification which has solely a single bird classification task. 
This is in line with the findings in \citep{cauzinille2025crossing} where the transfer from speech models varied depending on the layer chosen according to the final task and dataset. There is no single layer best across all datasets and taking all of them may be a better solution than training separate layer-wise models. The top-line fully fine-tuning offers the best performance, particularly for SSL, however at the cost of training a much larger model. 

This finding advocates for the adoption of multi-layer probing in bioacoustics benchmarks as it is the case for benchmarking speech transformers \citep{tsai2022superb}.
However, unlike SUPERB, where adapters are unnecessary because the embeddings have the same dimension, our approach requires adapters to project embeddings of varying shapes, while adding a significant number of parameters compared to the shallow probes.
This increased capacity for probes allows for more expressive transfer from large base models like the transformers, a fact observed in speech benchmarking \cite{zaiem2025speech}. 
In fact, we could confirm this finding with EfficientNet, a small CNN model (by the modern standards) that requires large adapters i.e. large number of parameters. In this case, using all layers in this network increased performance in accuracy by $0.09$ for BEANS classification and detection and with $0.02$ in mAP for BirdSet.

\noindent\textbf{Attention probes boost SSL models representations.}
Linear probes average or pool time-wise embeddings and apply a simple classification head. This loses the dependencies in these embeddings which may dampen performance. In a similar trend to multi-layer probing, we found attention probe heads to have superior transfer compared to linear probe heads, for all transformer models. We note large improvements for AVES (SSL), BEATs (SSL), EAT (SSL), NatureBEATs (SL) and less improvements for the other models. We note that the attention probe coupled with SSL models forms a powerful probing head, reported also in \cite{mahon2025robust}. This may be due to the way SSL models are trained: strongly modeling temporal dependencies in audio rather than learning species-wise inductive biases as in SL models. In other words, if your base model is a transformer trained in an SSL fashion it is better to pair it with another transformer to fully benefit from its representation.
More, attention probes did not improve with EfficientNet embeddings, and this may be due to the fact that the CNN operations in EfficientNet do not model time dependencies explicitly but consider the spectrogram as a 2D image.  

\noindent\textbf{Exploring the learned weights: SSL vs SL comparison.}
The multi-layer probing has the advantage that the learned weights for each layer can be explored for each dataset and base model. Considering the findings in the previous section, we want to gain further insights into what the probes are learning for the multi-layer case, whether the probes rely on fundamentally different representations when using SSL models, compared to using SL models.

\begin{figure}[h]
\centering
  \includegraphics[width=0.5\textwidth]{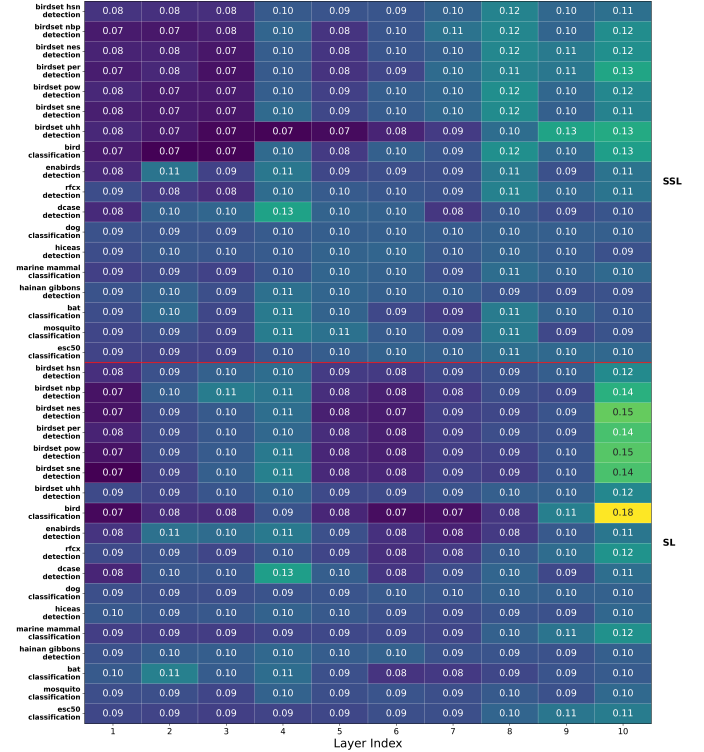}
  \caption{Learned layer weights across datasets when using multi-layer probing, averaged for SSL vs SL base models}
  \label{fig:weights}
\end{figure}

Given that most of the training data for the bioacoustics SL models comprises birds, we expect these models to specialize in bird species, particularly in the last layers i.e. where the models becomes more specialized to the task. At the same time, SSL models that we train on the same data do not model these bird species classification task explicitly.
We aggregate the results for the SSL and the SL models and we plot the weights learned for each dataset in BEANS and BirdSet in Figure \ref{fig:weights}. We group the datasets by their taxonomic proximity e.g. birds and mammals datasets are separated. We find that for birds species benchmarks, the first 10 rows, the probes rely more on the upper weights that become specialized in this task during the SL training, whereas the middle layers are more important when using SSL models. For mixed datasets as `dcase' and mammal taxa group the knowledge is more distributed across all the layers, whereas for bat individual id classification, the lower layers are more important. 

\section{CONCLUSION}
\label{sec:conclusion}
In this paper we propose a systematic evaluation of probing strategies for SSL and SL models on bioacoustic benchmarks. Our experiments show that using information from multiple layers works better than relying only on the final layer. We also found that pairing self-supervised transformers with attention-based probes is especially useful, since this setup makes the most of the temporal patterns learned by these models. Taken together, these results suggest that the common practice of using only a simple linear probe may underestimate how well an encoder actually performs. We therefore recommend that bioacoustic benchmarks include these richer probing setups to give a more accurate picture of model performance. We advise practitioners to use multi-layer probes for tasks other than bird species classification, and to use attention probe heads if their base models are SSL.  
 
\vfill\pagebreak

\bibliographystyle{IEEEbib}
\bibliography{bibliography}

\end{document}